\documentclass[12pt]{article}
\usepackage{graphicx} 
\usepackage{epsfig}
\usepackage{amsmath, amsfonts, amsthm,amssymb}
\usepackage{latexsym}
\usepackage{xcolor}
\usepackage{mathrsfs}
\usepackage{natbib}
\usepackage{hyperref}
\usepackage{appendix}
\usepackage{dsfont}
\usepackage[left=3cm,top=4cm,right=3cm,bottom=3cm]{geometry}
\RequirePackage[caption=false]{subfig}
\usepackage{lscape}
\usepackage{dsfont}
\usepackage{rotating}
\usepackage{ulem}

\usepackage{mathabx}
\usepackage{calligra}

\newcommand{\de}{\mathrm{d}}

\newcommand{\R}{\mathbb{R}}

\graphicspath{{./figs/}}

\begin{document}

\title{\LARGE \bf{Multiscale Topological Inference for Marked Point Processes via Euler Characteristic Envelopes} }
\maketitle
\begin{center}
{{\bf Matthias Eckardt$^{1}$} and {\bf Mehdi Moradi$^{2}$}}\\
\noindent $^{\text{1}}$ Chair of Statistics, Humboldt-Universit\"{a}t zu Berlin , Berlin, Germany\\
\noindent $^{\text{2}}$ Department of Mathematics and Mathematical Statistics, Ume\r{a} University, Ume\r{a}, Sweden
\end{center} 
\begin{abstract}
The statistical analysis of marked point processes requires disentangling complex spatial arrangements from attribute-dependent interactions. While classical summary statistics are effective for second-order dependencies, they frequently fail to capture higher-order topological structures and non-linear interactions between marks and space. In this work, we propose a novel multiscale topological inference framework for marked point processes by integrating mark-weighted filtrations with Euler Characteristic envelopes. We redefine the underlying metric space using an exponential mark-weighted distance, which modulates connectivity based on attribute similarity, effectively accelerating the merger of connected components among homophilic neighbors. To ensure rigorous statistical inference, we apply non-parametric global envelope tests to the resulting Euler Characteristic Curves, allowing for formal hypothesis testing against the null model of random labeling. Furthermore, we introduce a local decomposition of the topological signal via Z-scores at the critical filtration scale to identify and localize structural hubs and topological barriers. Systematic simulations across various scenarios demonstrate the framework's high specificity and sensitivity to attribute-space dependencies while remaining robust against purely geometric effects. This methodology provides a comprehensive and interpretable toolkit for identifying, quantifying, and localizing complex structural dependencies in marked spatial data, bridging the gap between topological data analysis and classical point process statistics.

\end{abstract}
{\it Keywords:  } 

\section{Introduction}

The statistical analysis of spatial point processes has historically focused on second-order summary statistics, such as the pair correlation function, Ripley's $K$-function, or nearest-neighbor distributions \citep{Eckardt:Moradi:currrent}. While these tools are indispensable for characterizing fundamental spatial tendencies like clustering, regularity, or inhibition, they are inherently limited to low-order, often isotropic interactions. Consequently, they frequently fail to capture higher-order topological structures—such as cycles, voids, or complex connectivity motifs—that arise from the intricate interplay between point locations and their associated attributes.

In many modern applications, from forest ecology to materials science, points are augmented with categorical or continuous marks. The challenge in analyzing such marked point processes lies in disentangling purely geometric aggregation from attribute-space coupling. Current state-of-the-art methods, including mark correlation functions  and inhomogeneous extensions \citep{moradi2026inhomogeneous}, typically evaluate mark dependencies conditionally on fixed Euclidean distances. This approach, however, neglects the potential for marks to fundamentally alter the connectivity landscape of the process, a limitation that hinders the detection of non-linear structural dependencies.

\begin{figure}[htbp]
    \centering
    \includegraphics[scale=.45]{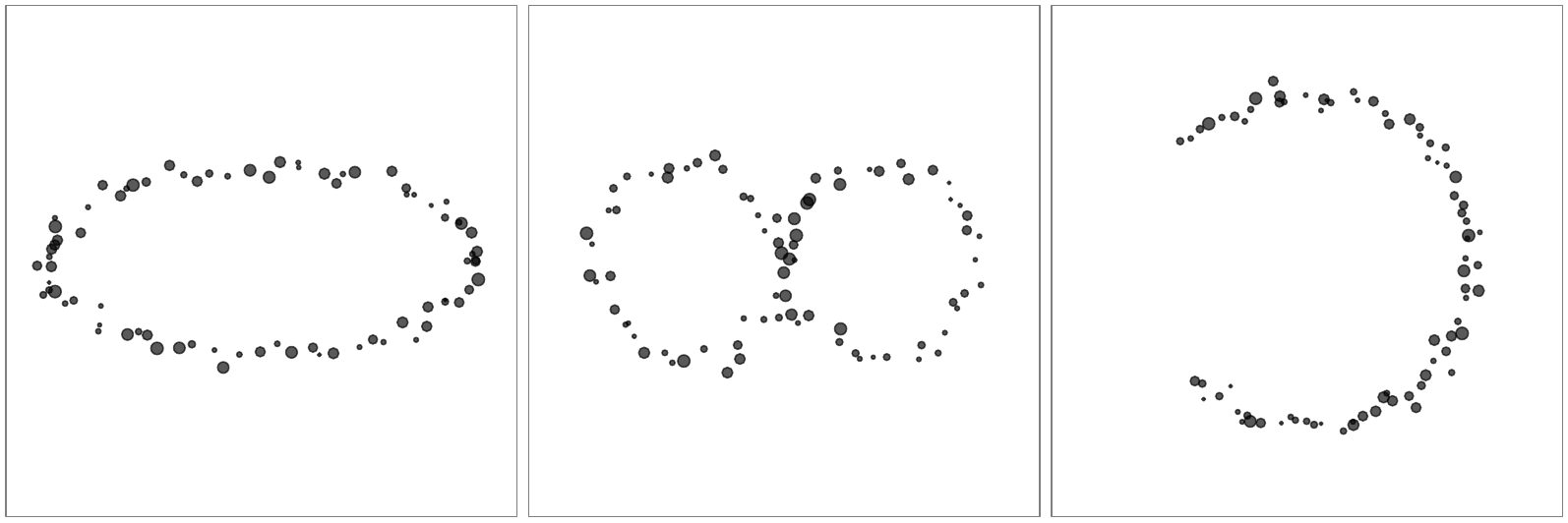}
    \caption{Structured marked point processes}
    \label{fig:intro}
\end{figure}

Topological data analysis (TDA), particularly persistent homology, offers a robust framework to address these limitations by tracking the evolution of $k$-dimensional structures across multiple scales \citep{Wasserman:2018:AnnualRev, Chazal:Bertrand:2021, carlsson2021topological, zoromodian}. Yet, standard TDA remains largely "blind" to mark information, as filtrations are traditionally constructed using Euclidean distances or density-based kernels. While Betti curves and persistence diagrams provide powerful descriptors of geometric connectivity, they do not intrinsically incorporate the rich information provided by marks, leaving a gap in the topological characterization of marked spatial point processes.

In this work, we propose a novel methodological framework that bridges this gap by redefining the underlying metric space through an \textit{exponential mark-weighted distance}. By scaling the Euclidean distance between any two points by the distance between the marks , we integrate attribute similarity directly into the birth-and-death process of topological features. This weighted filtration effectively accelerates the merger of components among homophilic neighbors while delaying connections between dissimilar points, thereby making the resulting Euler Characteristic Curves (ECC) highly sensitive to mark-space interactions.

Beyond mere description, we introduce a rigorous inferential framework by integrating the ECC with non-parametric Global Envelope Tests \citep{mari1, GETpack}. This allows for formal hypothesis testing against the null model of random labeling, a significant advancement over purely qualitative TDA summaries. Furthermore, we provide a local decomposition of the topological signal via $Z$-scores at the critical filtration scale $\epsilon_{crit}$, enabling the identification of specific structural hubs and outliers. By transforming attribute similarity into a topological connectivity driver, our method provides a comprehensive toolkit for identifying, localizing, and quantifying complex structural dependencies that remain invisible to classical point process statistics.

\section{Preliminaries}\label{sec:back}

Let $X = \{(x_i,m_i)\}_{i=1}^N, N<\infty$, be a marked point process in $\mathbb{R}^2 \times \mathbb{R}$, where $x_i \in \mathbb{R}^2$ denotes the location of $i$-th event and $m_i \in \mathbb{R}$ its associated real-valued mark. 
The intensity function $\lambda(u, m(u))$, $(u, m(u)) \in \R^2 \times \R$, and the second-order product density function $\lambda^{(2)}((u, m(u)), (v, m(v)))$ are, through the Campbell's formula, defined as
\begin{align*}
\mathbb{E}
\left[
\sum_{(x, m(x)) \in X} f(x, m(x))
\right]
&= 
\int
f(u, m(u))
\lambda(u, m(u)) 
\de u \
\nu(\de m(u)),
\end{align*}
and
\begin{align*}
&\mathbb{E}\!
\left[\sum_{(x, m(x)) \neq (y, m(y))}
f
\big(
(x, m(x)), (y, m(y))
\big)
\right]
\\
&=
\int
\int
f
\big(
(u, m(u)), (v, m(v))
\big)
\lambda^{(2)}
\big(
(u, m(u)), (v, m(v))
\big)
\de u \
\nu(\de m(u)) \
\de v \
\nu(\de m(v)),
\end{align*}
where $f$ is any non-negative measurable function and $\nu$ is the measure associated with the mark space. 
If the intensity function $\lambda$ is constant, the process is called homogeneous; otherwise it is said to be inhomogeneous.
The intensity function or product densities (of any order) describe only the expected number of points per unit volume in the product space or joint occurrences of points.
Consequently, they do not directly reveal the interaction structure among points, marks, or their joint dependence. 
To study such interactions, a commonly used summary statistic is the pair correlation function
\begin{align}\label{eq:pcf}
g
\big(
(u, m(u)), (v, m(v))
\big)
=
\frac{
\lambda^{(2)}
\big(
(u, m(u)), (v, m(v))
\big)
}{
\lambda(u, m(u))
\lambda(v, m(v))
},
\quad
(u, m(u)), (v, m(v)) \in \R^2 \times \R.
\end{align}
Given that for Poisson processes, 
$\lambda^{(2)}((u, m(u)), (v, m(v))) = \lambda(u, m(u)) \lambda(v, m(v))$, values of $g((u, m(u)), (v, m(v)))$ around one suggest complete spatial randomness,
values larger than one indicate clustering tendencies, and values less than one indicate inhibition between points.
Note that the pair correlation function \eqref{eq:pcf} characterizes interactions in the product space $\R^2 \times \R$.

Turning to spatial variability and interaction among marks, one needs to employ mark correlation functions.
These functions were originally defined for stationary point processes, i.e., processes whose distribution is translation invariant. However, \cite{moradi2026inhomogeneous} recently introduced a class of mark correlation functions for inhomogeneous marked point processes. Specifically, their construction is based on the class of second-order intensity-reweighted stationary processes, for which the pair correlation function  
$g((u, m(u)), (v, m(v)))$ depends only on the spatial distance $d(u,v)$. 
This framework encompasses the setting under which most summary statistics for inhomogeneous point processes are defined, including the inhomogeneous $K$-functions. 
Within this framework, \cite{moradi2026inhomogeneous} defined inhomogeneous mark correlation functions as the ratio of a mark-weighted inhomogeneous pair correlation function to the pair correlation function of the ground process. 
In other words, for a given test function $t_f:\mathbb{R}\times\mathbb{R}\to\mathbb{R}$, the unnormalised inhomogeneous mark correlation function is defined as
\begin{align}\label{eq:imcf}
c^{\mathrm{inhom}}_{t_f}(r)
&=
\frac{
\mathbb{E}
\left[
\sum_{(x,m(x)),(y,m(y)) \in X}^{\neq}
t_f(m(x),m(y)) 
\mathbf{1}\{ d(x,y)=r \}
/
\lambda(x)\lambda(y)
\right]
}{
\mathbb{E}
\left[
\sum_{(x,m(x)),(y,m(y)) \in X}^{\neq}
\mathbf{1}\{ d(x,y)=r \}
/
\lambda(x)\lambda(y)
\right]
}.
\end{align}
In practice, the condition $d(x,y)=r$ is interpreted through a small distance interval around $r$. One of the well-known mark correlation functions is the inhomogeneous mark variogram $\gamma^{\mathrm{inhom}}_{mm}$ for which $t_f(m_1,m_2) = 0.5 (m_1 - m_2)^2$. 
Under the assumption of mark independence, \eqref{eq:imcf} admits a closed-form expression, often referred to as the normalisation factor.
Values of \eqref{eq:imcf} are then compared with this factor to assess positive or negative mark associations or variations among points.
For instance, in the case of the inhomogeneous mark variogram $\gamma^{\mathrm{inhom}}_{mm}$, the normalisation factor equals the variance of the marks.
We refer to \citet{eckardt2024marked, moradi2026inhomogeneous} for further details on the use of different test functions and their interpretations.

\section{Multiscale Topological Inference}\label{sec:TDA}

Classical summary statistics for marked point processes, such as pair correlation functions or mark correlation functions introduced in Section~\ref{sec:back}, characterise mark-dependent interactions as functions of the interpoint distance $r$. 
While effective for describing second-order dependence among points, these statistics are intrinsically restricted to pairwise interactions. 
Consequently, spatial structures formed by groups of points, such as clusters of nearby points, loop-like structures surrounding empty regions, or larger voids that persist across multiple spatial scales, may not be detected. 
In contrast, topological data analysis (TDA) provides a complementary framework for quantifying the geometric and topological structure of point patterns in a coordinate-free and deformation-invariant manner \citep{carlsson2009topology}. 
By characterising connectivity and higher-order spatial relationships beyond pairwise distances, topological summaries capture aspects of multi-point spatial structure.

\subsection{TDA for unmarked point processes}

\begin{figure}
    \centering
    \includegraphics[scale=0.0489]{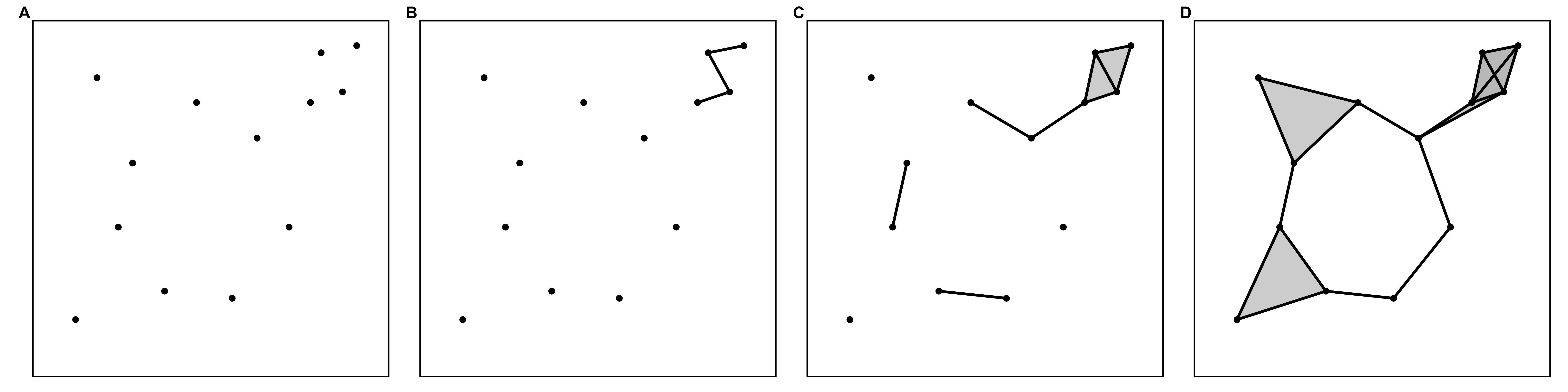}
    \caption{
Illustration of the Vietoris-Rips filtration.
Panels (A)--(D) show the simplicial complex $\mathcal{VR}(\mathbf{x},\epsilon)$ constructed for increasing values of the distance threshold $\epsilon$. 
For small $\epsilon$ (A), the complex consists primarily of isolated vertices. 
As $\epsilon$ increases (B), edges appear between nearby points, reflecting local proximity relations. 
At intermediate scales (C), larger connected structures emerge and loops enclosing empty regions become visible. 
For sufficiently large $\epsilon$ (D), higher-dimensional simplices fill these loops, illustrating the creation and disappearance of topological features that are subsequently tracked through persistent homology.}
\label{fig:ex}
\end{figure}

In the unmarked setting, given an observed point pattern $\mathbf{x}=\{x_i\}_{i=1}^n$ and a distance threshold $\epsilon>0$, one may construct geometric objects that encode proximity relations among the points. A widely used construction is the Vietoris-Rips complex $\mathcal{VR}(\mathbf{x},\epsilon)$ obtained by introducing a simplex whenever all pairwise Euclidean distances among its vertices do not exceed $\epsilon$. In particular, an edge is placed between any pair of points whose distance is at most $\epsilon$; whenever three points are mutually within distance $\epsilon$, the corresponding filled triangle is included; whenever four points are mutually within distance $\epsilon$, the corresponding tetrahedron is added. Higher-dimensional simplices arise analogously whenever all pairwise distances within a set of points do not exceed $\epsilon$.
Figure~\ref{fig:ex} illustrates the construction of the Vietoris-Rips filtration for an observed point pattern as the distance threshold $\epsilon$ increases.

The central combinatorial object underlying this construction is the simplicial complex. Formally, a simplicial complex $K$ is a finite collection of simplices satisfying two consistency conditions: every face of a simplex in $K$ must also belong to $K$, and the intersection of any two simplices is either empty or a common face. A $k$-simplex, denoted by $\sigma_k$, is the convex hull of $k+1$ affinely independent points. Concretely, $0$-simplices correspond to points, $1$-simplices to edges, $2$-simplices to filled triangles, and $3$-simplices to tetrahedra. Doing so, a simplicial complex extends the usual graph representation of a point pattern by encoding not only pairwise connections between points but also higher-order geometric relations among triples, quadruples, and larger groups of points. 
As the distance threshold $\epsilon$ increases, additional simplices are added to the complex. For small values of $\epsilon$, the complex consists only of isolated vertices corresponding to the observed points. As $\epsilon$ grows, edges appear between nearby points, triangles are formed once triples of points become mutually connected, and increasingly higher-dimensional simplices are introduced. This process generates a nested sequence of complexes
\begin{equation}
\mathcal{VR}(\mathbf{x},\epsilon_1)
\subseteq
\mathcal{VR}(\mathbf{x},\epsilon_2)
\subseteq
\cdots
\subseteq
\mathcal{VR}(\mathbf{x},\epsilon_{\max}),
\qquad 
\epsilon_1 < \epsilon_2 < \cdots < \epsilon_{\max},
\end{equation}
known as a simplicial filtration. 
The filtration provides a multiscale representation of the spatial structure of the point pattern. Topological features such as connected components, loops, and enclosed voids appear at certain values of $\epsilon$ and disappear at larger thresholds when they merge or become filled. Tracking the birth and death of such features across the filtration constitutes the central idea of persistent homology. 
Several alternative simplicial constructions based on the same multiscale principle have been proposed in the literature, including the \v{C}ech complex, the $\alpha$-complex, the graph-induced complex \citep{DEY2015575}, the witness complex \citep{desilva:witness}, and the $\alpha\beta$-witness complex \citep{10.1007/978-3-540-73951-7_34}. In the present work, however, we focus on the Vietoris-Rips construction.

\subsection{TDA for marked point processes}

A fundamental limitation of the classical construction is that the filtration depends exclusively on spatial locations: the associated marks do not influence when two points become topologically connected. Consequently, for a marked point pattern $\mathbf{x}=\{(x_i,m_{x_i})\}_{i=1}^n$, structural patterns arising from the joint spatial–mark configuration, such as clustering among points with similar mark values, remain invisible unless they coincide with purely geometric proximity. 
To illustrate, consider a configuration in which two types of points, distinguished by their marks, are spatially intermingled. Suppose that points with large and small marks form two distinct attribute groups but are interleaved in space. Distances between points from different groups may then be smaller than distances within the same group. Consequently, the standard Vietoris–Rips filtration connects points across the two groups at small values of $\epsilon$, rapidly merging them into a single component and thereby obscuring the mark-driven separation. 
Conversely, two points that are spatially distant but have nearly identical marks may belong to a meaningful mark-homogeneous cluster. Since the classical filtration depends only on spatial distance, such relationships cannot be detected at any intermediate scale. These examples illustrate that filtrations based solely on spatial proximity may fail to capture important multi-point structure arising from the joint spatial–mark configuration. 
To address this limitation, we replace the Euclidean distance with the  mark-weighted distance $d_M$
\begin{equation}
d_M\!\left((x,m_x),(y,m_y)\right)
=
d(x,y)\exp\!\left(|m_x-m_y|\right),
\label{eq:weighted_dist}
\end{equation}
which couples spatial separation with mark dissimilarity. Using this distance, we define the mark-weighted Vietoris–Rips complex
\begin{equation}
\mathcal{VR}_m(\mathbf{x},\epsilon)
=
\left\{
\sigma \subseteq \mathbf{x} \;\middle|\;
d_M(u,v) \le \epsilon
\text{ for all } u,v \in \sigma
\right\}.
\label{eq:vr_complex}
\end{equation}
A subset $\sigma \subseteq \mathbf{x}$ with $|\sigma|=k+1$ corresponds to a $k$-simplex, consistent with the notation introduced earlier. 
The distance $d_M$ acts as a joint dissimilarity measure combining spatial proximity and mark similarity through exponential weighting. If two points are spatially close and carry similar marks so that $|m_i-m_j|\approx 0$, then $d_M\approx d(x_i,x_j)$ and the pair enters the filtration at essentially the same scale as under the standard Euclidean construction. If two points are spatially close but have strongly contrasting marks, the exponential factor inflates their effective distance, delaying their topological connection until a larger value of $\epsilon$. Conversely, if two points are spatially far apart but have nearly identical marks, the exponential factor remains close to one and their separation is still dominated by the Euclidean distance, so they enter the filtration only at a large scale.

For each $\epsilon$, the filtration generates a simplicial complex $\mathcal{VR}_m(\mathbf{x},\epsilon)$. To quantify the structural information encoded in these complexes, TDA employs homology, which provides a systematic framework for identifying and counting topological features of different dimensions. These features are characterised by the homology groups $H_k(\mathcal{VR}_m(\mathbf{x},\epsilon))$, where $H_k$ denotes the $k$-th homology group whose rank counts $k$-dimensional topological features such as connected components $(k=0)$, loops $(k=1)$, and higher-dimensional voids. 
The ranks of these groups, known as the Betti numbers $\beta_k$, quantify how many such features are present. Formally,
\begin{equation}
\beta_k(\epsilon)
=
\mathrm{rank}\!\left(
H_k\!\left(\mathcal{VR}_m(\mathbf{x},\epsilon)\right)
\right),
\label{eq:betti}
\end{equation}
which gives the number of independent topological features of dimension $k$ at scale $\epsilon$. For interpretational purposes, $\beta_0(\epsilon)$ counts the number of connected components in $\mathcal{VR}_m(\mathbf{x},\epsilon)$, while $\beta_1(\epsilon)$ counts the number of independent loops formed by chains of edges. 
As the threshold $\epsilon$ increases along the filtration, isolated points and small components progressively merge into larger connected groups, leading to a decrease in $\beta_0$, while loops may appear and later disappear as triangles fill them in. Under the mark-weighted distance \eqref{eq:weighted_dist}, these quantities admit a natural interpretation for marked point patterns. In particular, $\beta_0(\epsilon)$ measures the number of spatial clusters of points that are both close in space and similar in their marks. Since connectivity now requires both spatial proximity and mark similarity, spatially adjacent groups of points with strongly contrasting marks remain topologically separated until $\epsilon$ becomes large enough to overcome the exponential mark penalty. 
Similarly, $\beta_1(\epsilon)$ counts loops formed by chains of points connected through the mark-weighted proximity rule \eqref{eq:weighted_dist}.
Such loops may arise when chains of spatially close and mark-similar points form ring-like structures that separate regions with contrasting mark values. While $\beta_0(\epsilon)$ and $\beta_1(\epsilon)$ each provide informative multiscale summaries, it is often statistically and computationally convenient to combine the full topological profile into a single functional descriptor. Here we use the Euler characteristic curve (ECC) \citep{edelsbrunner2010computational}, obtained by
\begin{equation}
\chi(\epsilon)
=
\sum_{k \geq 0} (-1)^k \beta_k(\epsilon).
\label{eq:euler}
\end{equation}

For interpretational purposes, when only connected clusters and ring-like structures are present the Euler characteristic reduces to $\chi(\epsilon)=\beta_0(\epsilon)-\beta_1(\epsilon)$. 
Here, $\beta_0(\epsilon)$ counts the number of connected components in the complex, while $\beta_1(\epsilon)$ counts the number of loop-like configurations formed by chains of edges surrounding empty regions. 
The ECC therefore reflects the balance between connected components and loops as the filtration scale $\epsilon$ increases. 
Large positive values of $\chi(\epsilon)$ indicate that connected components dominate over loops, corresponding to a fragmented or dispersed spatial configuration of points, whereas small or negative values indicate a relative abundance of loops, corresponding to ring-like or locally enclosed structures. 
As a functional summary, the ECC plays a role analogous to classical point process summary statistics, summarising the multiscale topological structure of the observed point pattern as a function of the filtration parameter $\epsilon$. 
More generally, the Euler characteristic combines contributions from topological features of different dimensions according to \eqref{eq:euler}. Higher-order Betti numbers typically contribute little, so the behaviour of $\chi(\epsilon)$ is largely driven by the interplay between $\beta_0(\epsilon)$ and $\beta_1(\epsilon)$. 
The Euler characteristic is a fundamental invariant from integral geometry and one of the Minkowski functionals characterised by Hadwiger's theorem. 
Evaluating this quantity along the filtration produces the ECC, which summarises the evolution of topological structure across scales. 
Since the filtration is defined using the mark-weighted distance $d_M$, the resulting ECC reflects joint spatial-mark structure. 
Consequently, deviations of $\chi(\epsilon)$ from the behaviour expected under random labelling indicate dependence between spatial configuration and marks, rather than effects arising purely from spatial structure or from the distribution of marks alone.

\subsubsection{Global envelop tests}

To assess whether the observed coupling between marks and spatial structure departs from random labelling, we consider the null hypothesis that the marks are randomly reassigned across the fixed spatial locations. Under this null model, the spatial configuration remains unchanged while the marks are permuted among the observed points, implying that any association between marks and spatial positions arises purely by chance. 
A reference distribution for the Euler characteristic curve $\chi(\epsilon)$ is generated by performing $s$ Monte Carlo permutations of the marks, producing simulated curves $\{\chi_{\mathrm{sim},j}(\epsilon)\}_{j=1}^s$. These simulated curves represent the variability of the ECC expected under the random labelling hypothesis. 
Statistical significance is assessed using a global envelope test \citep{mari1, GETpack, myllymaki2019get} constructed from the simulated ensemble. Specifically, the observed curve $\chi_{\mathrm{obs}}(\epsilon)$ is compared with the simulated curves across the full range of $\epsilon$. If the observed curve falls outside the global envelope at any scale, the null hypothesis of random labelling is rejected. 
Additionally, the resulting empirical $p$-value provides an interpretable measure of whether the observed topological signature of the marked point pattern is consistent with the random labelling null model. 
If $\chi_{\mathrm{obs}}(\epsilon)$ falls below the lower boundary of the envelope, topological connections in the observed pattern form earlier than expected under the null model. This corresponds to a faster merging of connected components (lower $\beta_0$), the formation of more loops (higher $\beta_1$), or both. Under the mark-weighted distance \eqref{eq:weighted_dist}, such behaviour arises when spatially proximate points tend to carry similar marks, so that the exponential penalty remains close to unity and connections form at smaller values of $\epsilon$. If the observed ECC lies below the lower envelope across a range of scales, this indicates that topological connections form earlier than expected under random labelling and provides a topological signature of positive spatial autocorrelation of the marks.  
Conversely, if $\chi_{\mathrm{obs}}(\epsilon)$ lies above the upper boundary of the envelope, neighbouring points tend to carry contrasting marks. In this case the exponential weighting inflates the effective distances between nearby points, delaying component mergers and maintaining a larger number of connected components than expected under random labelling. This behaviour corresponds to mark segregation or repulsion, where local mark contrasts act as a barrier to topological connectivity.

\subsubsection{Individual Z-scores}

The global envelope test identifies whether and at which range of scales significant mark–space coupling is present, but it does not directly indicate which locations in the point pattern are responsible for the observed signal. To connect global significance with spatial interpretation, we identify a critical filtration scale
\begin{equation}
\epsilon_{\mathrm{crit}}
=
\arg\max_{\epsilon}
\left|
\mathrm{rank}_{\mathrm{sim}}
\!\left(
\chi_{\mathrm{obs}}(\epsilon)
\right)
-
\mathrm{median\;rank}
\right|,
\end{equation}
that is, the scale at which the observed ECC deviates most strongly from the distribution generated under random labelling. This scale corresponds to the point along the filtration where the coupling between marks and spatial structure is most pronounced. 
Fixing the filtration at $\epsilon_{\mathrm{crit}}$, we then examine the local contribution of individual points to the connectivity structure of the complex. For each point, a local score measuring its influence on $\chi(\epsilon_{\mathrm{crit}})$ is computed, yielding a spatial map of structural importance that highlights the points driving the global deviation from the null model. 
Points with large positive scores act as connectivity hubs: their marks are highly consistent with those of nearby points, reducing the exponential penalty in \eqref{eq:weighted_dist} and promoting early topological connections. Conversely, points with large negative scores act as structural outliers: their marks differ significantly from those of their neighbours, inflating effective distances and delaying the merging of components. 
Together, the global envelope test and the local contribution analysis provide a two-level inferential framework, wherein the global test establishes whether the observed topological signature is inconsistent with random labelling, while the local analysis identifies the specific spatial locations responsible for this deviation.

\section{Simulation Study}

We evaluate the performance of the proposed framework across nine simulation scenarios, each designed to isolate a specific combination of spatial and mark structure including CSR, clustered, and regular point and random, positively, and negatively correlated mark structures. All scenarios are conducted within the observation window $W = [0, 10]^2$ with $n = 80$ points, and global significance is assessed via $N = 999$ Monte Carlo permutations. The nine scenarios arise from a factorial design combining three spatial point patterns with three mark structures, allowing us to systematically assess which combinations the ECC can detect. 

The results from the simulation showing the marks point patterns, the ECC curves and the Z values for the nine different scenarios are depicted in Figures \ref{fig:CSR} to \ref{fig:Regular}. For interpretation, recall  that the ECC is defined as $\chi(\varepsilon) = \beta_0(\varepsilon) - \beta_1(\varepsilon)$, where $\beta_0$ counts connected components and $\beta_1$ counts independent cycles at filtration radius $\varepsilon$. Starting at $\varepsilon = 0$ all $n = 80$ points remain unconnected singletons with $\chi(0) = 80$. As $\varepsilon$ increases, components start to merge yielding a  reduction in $\beta_0$ while loops form and fill resulting first in an increase and then a decrease in $\beta_1$. Therefore, the ECC declines, potentially, dipping below zero and eventually stabilising at $\chi = 1$. The blue dashed curve represent  the observed ECC excluding any mark information and is tested against the CSR envelope. In constrast, the red solid curve depicts the mark-weighted ECC that is tested against a random-labelling envelope. Taking the Betti numbers into account, a curve leaving its envelope upward at small $\varepsilon$ highlights early merging, i.e.,  clustered structure or similar neighbours reducing effective distances whereas a curve leaving downward at larger $\varepsilon$ signals delayed merging, indicating regular spacing or dissimilar neighbours inflating effective distances.

\subsection{Simulated mark structures}

For the random mark scenario, the marks are drawn independently of the point locations,
\begin{equation}
    m(x_i) \overset{\mathrm{iid}}{\sim} \mathcal{U}(0, 8),
\end{equation}
establishing a null baseline with no mark-space coupling. The common range $[0, 8]$ yielding $\exp(8) \approx 2981$ for the most dissimilar pairs is chosen to match the structured mark scenarios below, ensuring that the permutation envelope is scale-consistent across all three mark structures.

To introduce positive correlation, marks are generated to exhibit positive spatial autocorrelation so that nearby points tend to carry similar mark values. For the CSR pattern, marks are obtained from a spatially smooth Gaussian random field. Specifically, let $\mathbf{z} \sim \mathcal{N}(\mathbf{0}, K_{gg})$ be a realisation of a zero-mean Gaussian process on a regular $20 \times 20$ grid covering $W$, with covariance kernel
\begin{equation}
    [K_{gg}]_{jk} = \exp\!\left(-\frac{\|\mathbf{g}_j - \mathbf{g}_k\|}{\rho}\right)
    + \nu \,\mathbf{1}_{j=k},
    \qquad \rho = 9,\; \nu = 0.001,
\end{equation}
where $\mathbf{g}_j$ denotes the $j$-th grid node. The range parameter $\rho = 9$ is chosen to  ensure that the GRF varies smoothly across the entire domain and that virtually all nearest-neighbour pairs share similar mark values. The nugget $\nu = 0.001$ is set close to zero to suppress local noise and enforce a near-deterministic smooth field. The mark at each point $x_i$ is obtained by kriging interpolation,
\begin{equation}
    m(x_i) = \mathbf{k}(x_i)^\top K_{gg}^{-1} \mathbf{z},
\end{equation}
where $[\mathbf{k}(x_i)]_j = \exp(-\|x_i - \mathbf{g}_j\| / \rho)$, and the resulting values are linearly rescaled to $[0, 8]$. For the clustered pattern, marks are instead assigned via cluster-specific means,
\begin{equation}
    m(x_i) = \mu_{c(i)} + \eta_i, \qquad \eta_i \overset{\mathrm{iid}}{\sim} \mathcal{N}(0,\, 0.01^2),
\end{equation}
where $c(i)$ denotes the cluster membership of point $x_i$ and the cluster means $\mu_{c(i)}$ alternate between $0$ and $8$ across clusters, yielding $\exp(8) \approx 2981$ for inter-cluster pairs, while the within-cluster noise $\sigma = 0.01$ keeps same-cluster marks nearly identical. So  spatially proximate points within the same cluster carry nearly identical marks while points from neighbouring clusters are maximally dissimilar. 
For the regular pattern, marks follow a deterministic periodic field,
\begin{equation}
    m(x_i) = 4\sin\!\left(\frac{12\pi\, x_{i,1}}{4.5}\right)\cos\!\left(\frac{2\pi\, x_{i,2}}{2.5}\right) + 4,
\end{equation}
clipped to $[0, 8]$ with small additive noise $\eta_i \overset{\mathrm{iid}}{\sim} \mathcal{N}(0,\, 0.05^2)$. The period of $2.5$ units in the $x_1$-direction and $2.5$ units in the $x_2$-direction   ensuring that each half-period contains several points with similar marks separated from points in the adjacent half-period by large mark differences. The amplitude of $4$ centres the field on $[0, 8]$ after the additive shift of $4$. The noise standard deviation $\sigma = 0.05$ is small relative to the amplitude, so the spatial structure dominates over local fluctuations. This induces strong periodic spatial autocorrelation that is clearly detectable under permutation.

In the final mark setting, marks are constructed to induce negative spatial autocorrelation via a $4 \times 4$ grid partition of $W$. Precisely, assigning a  block index $b$ with 
\begin{equation}
    b(x_i) = \left\lfloor \frac{x_{i,1}}{2.5} \right\rfloor
            + \left\lfloor \frac{x_{i,2}}{2.5} \right\rfloor
\end{equation}
to each point $x_i$, the mark are defined as
\begin{equation}
    m(x_i) =
    \begin{cases}
        0 + \eta_i & \text{if } b(x_i) \text{ is even,} \\
        8 + \eta_i & \text{if } b(x_i) \text{ is odd,}
    \end{cases}
    \qquad \eta_i \overset{\mathrm{iid}}{\sim} \mathcal{N}(0,\, 0.01^2).
\end{equation}
This forces virtually all nearest-neighbour pairs to cross a cell boundary, so that $|m(x_i) - m(x_j)| \approx 8$ for close pairs, yielding a mark-weighted distance inflation factor of $\exp(8) \approx 2981$. The cell width of $2.5$ units is larger than typical nearest-neighbour distances   but small enough that most pairs within a neighbourhood of radius $\delta = 2$ cross a cell boundary. This ensures that virtually all nearest-neighbour pairs satisfy $|m(x_i) - m(x_j)| \approx 8$, yielding a mark-weighted distance inflation factor of $\exp(8) \approx 2981$. The noise standard deviation $\sigma = 0.01$ is negligible relative to the mark contrast of $8$, so the grid structure is effectively deterministic. 

When marks are spatially independent, no mark-space coupling is present regardless of the point geometry. These three scenarios establish the specificity of the test. When marks exhibit positive spatial autocorrelation, the mark-weighted distance $d_M$ contracts distances between similar neighbours and expands distances between dissimilar ones, and the ECC is expected to deviate from the permutation envelope. This effect is strongest for the cluster pattern, where spatial proximity and mark similarity are simultaneously enforced within clusters. When marks are negatively correlated, adjacent points are systematically dissimilar, creating topological barriers in the filtration that delay component mergers across all three spatial configurations.

\subsection{Scenario I: CSR}

As initial setting points are drawn from a homogeneous Poisson process on $W$,
\begin{equation}
    X \sim \mathrm{HPP}(\lambda),
\end{equation}
establishing a baseline of complete spatial randomness (CSR) with no geometric clustering.  Focussing on the spatial structure, $n_{\mathrm{sim}} = 999$ realisations are simulated from an inhomogeneous Poisson process whose intensity is estimated from the observed pattern via kernel density estimation with the CvL bandwidth selector, and the ECC is computed without marks on each simulated pattern. For the mark structure, the observed marks are permuted uniformly at random across the fixed point locations $n_{\mathrm{sim}} = 999$ times (random labelling), and the mark-weighted ECC is computed on each permutation.  

\begin{figure}[htbp]
    \centering
    \includegraphics[scale=.5]{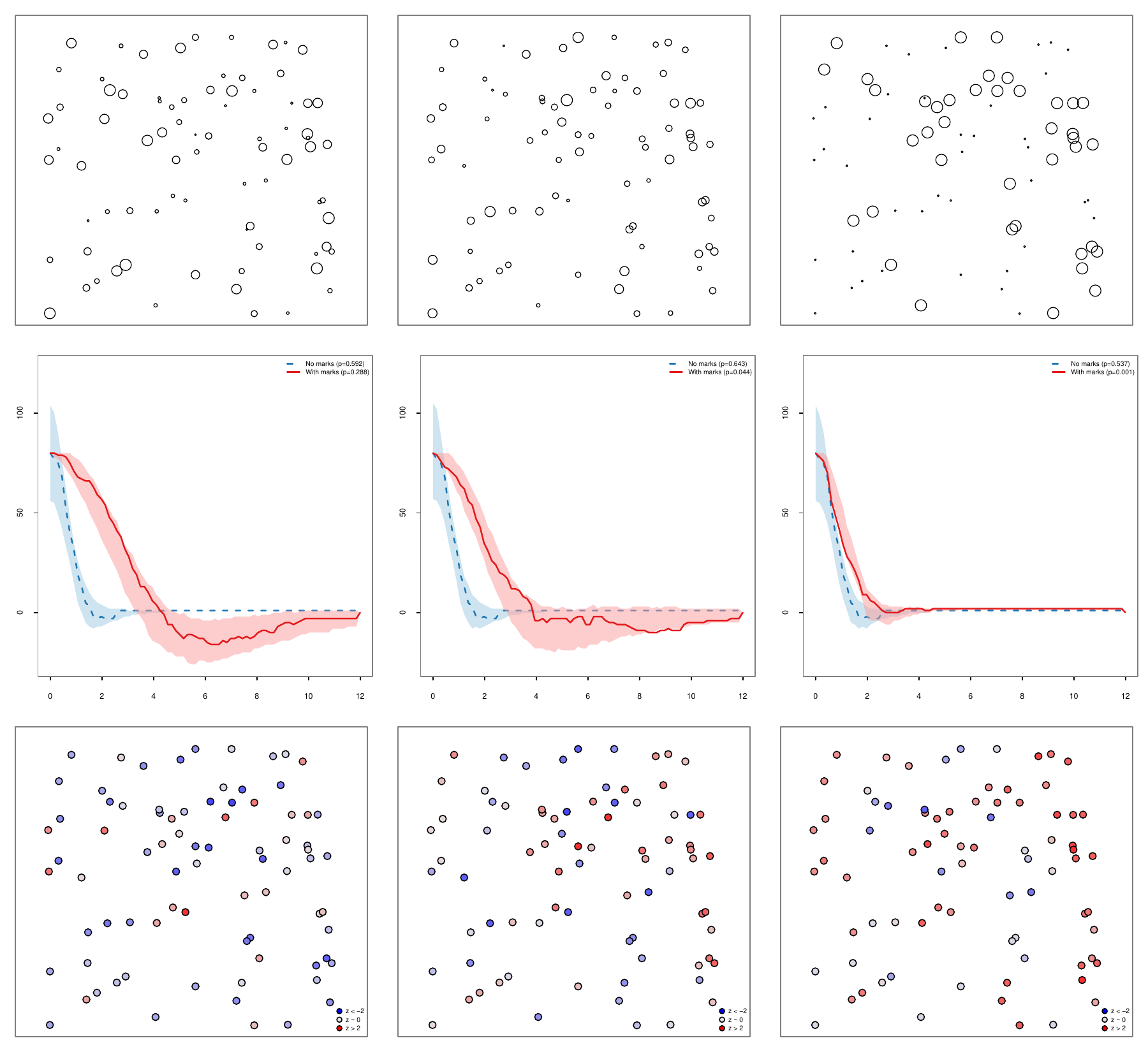}
    \caption{Scenario I: Simulated CRS point pattern with independent, positively correlated and negatively correlated mark from left to right. Simulated point pattern with corresponding mark structures  (top row), Euler characteristic curves and envelopes computed over 999 simulation for unmarked (blue) and marked (red) point patterns (central row), and computed $Z$-values (bottom row)}
    \label{fig:CSR}
\end{figure}

For the CSR point structures depicted in 
CRS Figure \ref{fig:CSR} the blue curves in all three mark scenarios remains inside the envelope, confirming that the point pattern is consistent with complete spatial randomness, i.e. the reference configuration.
The ECC of CSR pattern with random marks shown in the left central panel $p_{\text{red}} = 0.288$ indicating that the marks impose no systematic structure on the mark-weighted distances. The red curve is indistinguishable from its permutation envelope. Turning to the CSR pattern with positively correlated marks (central plot)    $p_{\text{red}} = 0.044$. The very smooth GRF ($\rho = 9$) assigns similar mark values to spatially proximate points such that the mark weight $\exp(|\Delta m|) \approx 1$ for close pairs while distant pairs receive inflated distances, causing components to merge slightly earlier than under random labelling. In constrast, for the negatively correlated mark setting with  $p_{\text{red}} = 0.001$, the $4 \times 4$ grid  assigns opposing values $\{0, 8\}$ to virtually all nearest-neighbour pairs, yielding $\exp(|\Delta m|) \approx \exp(8) \approx 2981$ for close pairs. All short-range merges are massively delayed yielding a departure of the red curve departs downward at larger $\varepsilon$.
 
\subsection{Scenario II: Clustering }

\begin{figure}[htbp]
    \centering
    \includegraphics[scale=.5]{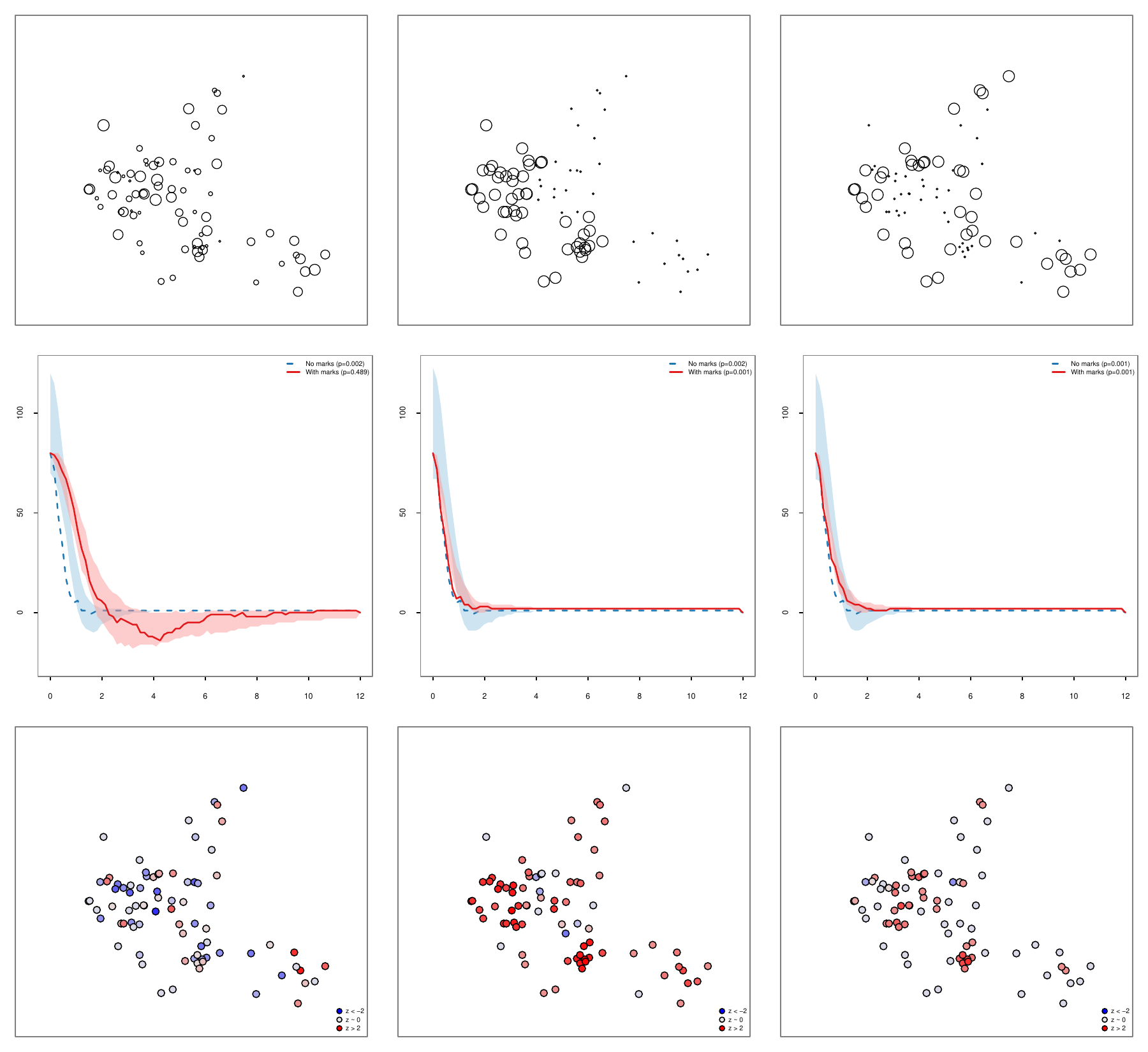}
    \caption{Scenario II: Simulated Thomas cluster process with independent, positively correlated and negatively correlated mark from left to right. Simulated point pattern with corresponding mark structures (top row), Euler characteristic curves and envelopes computed over 999 simulation for unmarked (blue) and marked (red) point patterns (central row), and computed $Z$-values (bottom row)}
    \label{fig:Thomas}
\end{figure}

For the clustered point configuration, locations are generated from a Thomas cluster process, where a parent process $\Phi_p \sim \mathrm{HPP}(\kappa)$ produces offspring clusters, with each parent $y \in \Phi_p$ generating a Poisson$(\mu)$ number of offspring distributed as
\begin{equation}
    x_i \mid y \;\sim\; \mathcal{N}_2(y,\, \sigma^2 I_2), \qquad \sigma = 0.7,
\end{equation}
inducing spatial aggregation. All blue ECC curves of Figure \ref{fig:Thomas} depart from the CSR envelopes at small $\varepsilon$. Within-cluster points are spatially closer and merge already at small radii such that $\beta_0$ drops far faster than expected under CSR. This corresponds to  the canonical topological signature of clustering.

The ECC for the random marks ($p_{\text{red}} = 0.489$) show that despite the strong spatial clustering of the points the i.i.d.\ marks impose no additional structure on the mark-weighted distances and he red curve lies firmly within its envelope. In contrast, for the clustered point pattern with positively correlated marks ($p_{\text{red}} = 0.001$) the cluster-specific means alternate between $0$ and $8$ such that the points within the same cluster carry nearly identical marks ($|\Delta m| \approx 0$, $\exp(|\Delta m|) \approx 1$) while inter-cluster distances are inflated by a factor of $\exp(8)$. As a result, within-cluster merges occur even earlier than expected and the red curve departs strongly upward at small scales $\varepsilon$. Both point and mark structure act in the same direction. Turning to the negatively correlated marks ($p_{\text{red}} = 0.001$), the grid imposes large mark contrasts even within clusters, inflating all short-range distances by $\exp(8)$. By this, all within-cluster and between-cluster merges are severely delayed and the red curve departs downward at larger $\varepsilon$. So both, the point and the mark structure act in opposite directions. Clustering accelerates the filtration while negatively correlated marks retard it, and the mark effect dominates.

\subsection{Scenario III: Inhibition}

\begin{figure}[htbp]
    \centering
    \includegraphics[scale=.5]{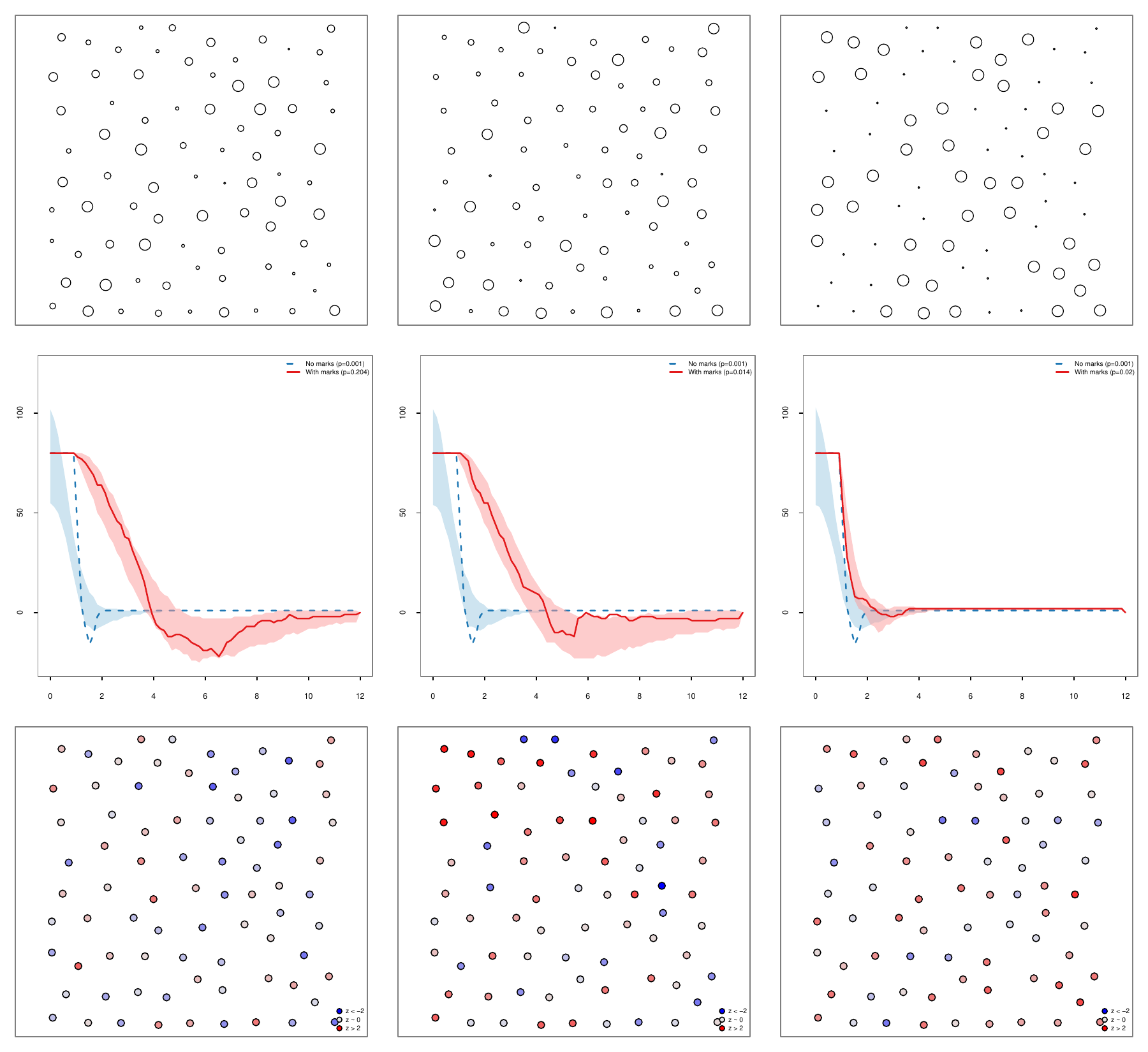}
    \caption{Scenario III: Simulated inhibition process with independent, positively correlated and negatively correlated mark from left to right. Simulated point pattern with corresponding mark structures (top row), Euler characteristic curves and envelopes computed over 999 simulation for unmarked (blue) and marked (red) point patterns (central row), and computed $Z$-values (bottom row)}
    \label{fig:Regular}
\end{figure}

Point locations are generated from a hard-core inhibition process with minimum separation distance $\delta = 0.9$, producing a spatially regular configuration in which points are systematically repelled from one another. The results for the inhibition process depicted in Figure \ref{fig:Regular}
highlight that all blue curve departs downward from the CSR envelope at larger values of $\varepsilon$. The hard-core inhibition distance $\delta = 0.9$ prevents any two points from being closer than $0.9$ units, so no early merges occur and $\beta_0$ decreases more slowly than under CSR. This delayed onset of connectivity is the topological signature of spatial regularity. Turning to the red curves, we found that the i.i.d.\ marks introduce no systematic mark-distance coupling and the ECC  stays within its envelope, confirming specificity under inhibition  ($p_{\text{red}} = 0.204$). For the positively correlated marks ($p_{\text{red}} = 0.014$), the periodic sinusoidal field assigns similar marks to spatially proximate points such that close pairs have $\exp(|\Delta m|) \approx 1$ and  effective distances are shorter than under expected with merges occur slightly earlier. As a result, the red curve departs upward, counteracting the downward departure of the blue curve such that the point and the mark structure act in opposite directions. For the inhibition process with negatively correlated marks ($p_{\text{red}} = 0.020$), the grid inflates all short-range distances by $\exp(8)$ yielding further delaying an already slow filtration. This results in a downward departure of the red curve, reinforcing the downward departure of the blue curve. Hence, both point and mark structure act in the same direction.
 
\subsection{Monte Carlo characterisation of the mark-weighted ECC}
 
To complement the global envelope tests presented above, we conduct a Monte Carlo characterisation of the mark-weighted Euler characteristic curve across all nine simulation scenarios. Rather than testing a single observed pattern against a null envelope, we simulate $B = 1000$ independent realisations of each scenario, compute both the plain (unweighted) and the mark-weighted ECC for every replicate, and summarise the resulting empirical distributions via their pointwise median and $2.5\%$/$97.5\%$ quantile bands. This yields, for each scenario, a direct characterisation of the typical behaviour of both curves under the corresponding joint point-and-mark process, without reference to a fixed observed pattern or a hypothesis test.

\begin{figure}[htbp]
    \centering
    \includegraphics[scale=.5]{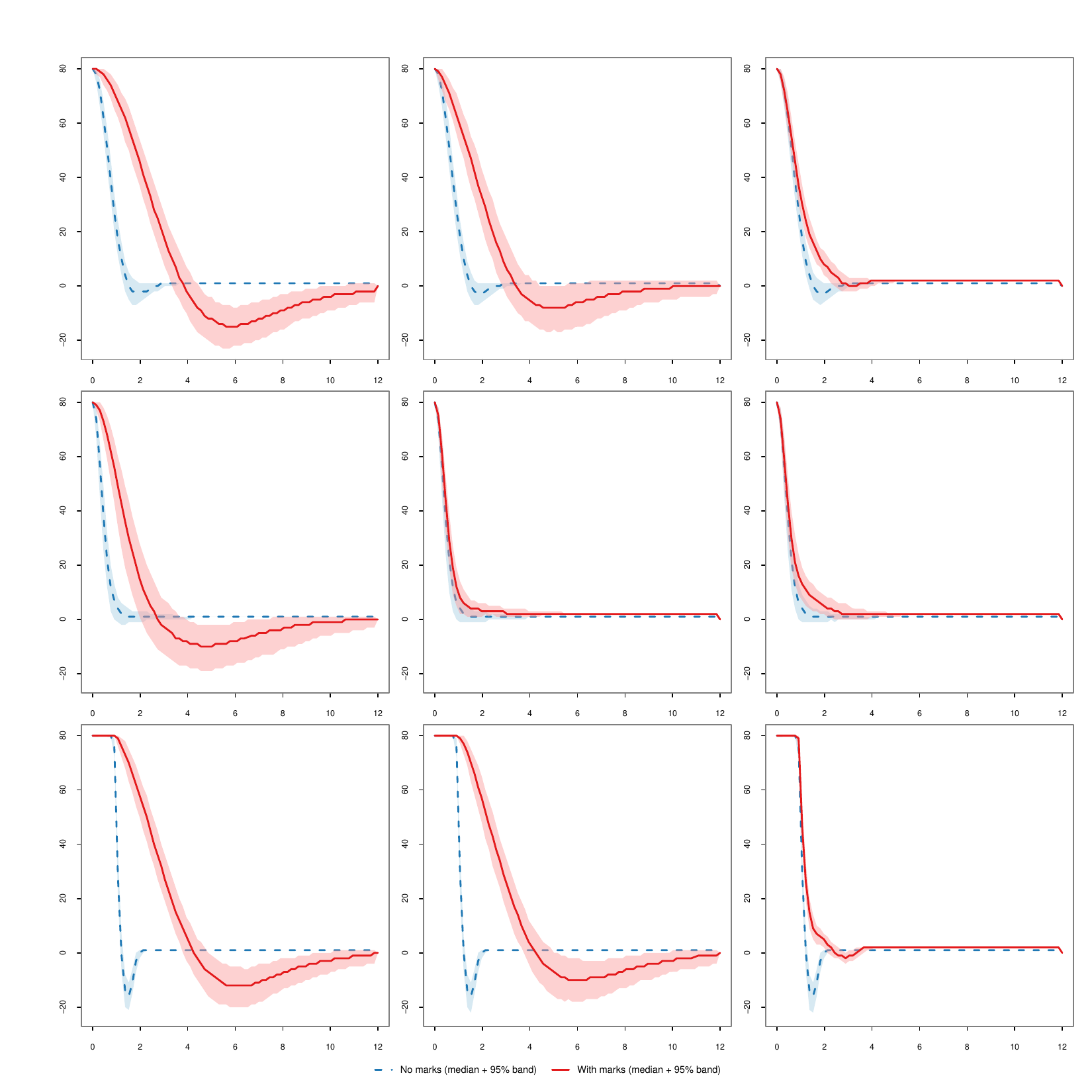}
    \caption{Results of the Monte Carlo Simulation for all nine simulated point patterns with $B = 1000$ replicates within each scenario. Mean EEC curves for unmark (blue) and mark (red) patterns and corresponding envelopes over $B = 1000$ replication for simulated CSR, Thomas cluster and hard-core inhibition (top to bottom) with independent, positively and negatively correlated marks (left to right)}
    \label{fig:mc_char}
\end{figure}
For each of the $B = 1000$ replicates within a scenario, point patterns are simulated from the corresponding spatial process (CSR, Thomas cluster, or hard-core inhibition with $\delta = 0.9$), with marks drawn from the corresponding mark model (random, positively correlated, or negatively correlated), and both the plain ECC
\begin{equation}
    \chi_{\mathrm{plain}}(\varepsilon) = \beta_0(\varepsilon) - \beta_1(\varepsilon)
\end{equation}
based on Euclidean distances, and the mark-weighted ECC
\begin{equation}
    \chi_{\mathrm{marked}}(\varepsilon) = \beta_0^M(\varepsilon) - \beta_1^M(\varepsilon)
\end{equation}
based on mark-weighted distances $d_M(x_i, x_j) = \|x_i - x_j\| \exp(|m(x_i) - m(x_j)|)$, are computed via the Vietoris--Rips filtration.

The nine panels of Figure~\ref{fig:mc_char} display the pointwise median (solid or dashed line) and $95\%$ band (shaded region) for the plain ECC (blue, dashed) and the mark-weighted ECC (red, solid) under each combination of spatial and mark structure. The results are consistent with the theoretical expectations derived above and confirm the behaviour observed in the single-pattern envelope tests.
For the effect of spatial structure (blue bands across rows), the plain ECC bands differ systematically across the three spatial configurations. Under CSR the blue band reflects the typical connectivity growth of a homogeneous Poisson process. Under the Thomas cluster process the blue band shifts upward and to the left as the within-cluster proximity of points causes components to merge at smaller radii with $\beta_0$ decreasing faster and the ECC remains elevated at small $\varepsilon$ before dropping sharply. Under hard-core inhibition the blue band shifts downward and to the right.  The minimum separation of $\delta = 0.9$ delays all component merges such that the ECC decreases more slowly and the band is positioned below the CSR reference. These shifts are consistent across all three columns, confirming that the plain ECC captures the spatial point structure independently of the mark model.

Turning to the effects of mark structure (red bands across columns), we observed that the mark-weighted ECC bands vary systematically across the three mark structures. Under i.i.d.\ marks the red band closely overlaps with the blue band in all three rows as no systematic distortion of the effective distances is introduced and  the mark-weighted filtration behaves similarly to the unmarked one. Under positively correlated marks the red band shifts upward relative to the random-mark case as spatially proximate points carry similar marks with $\exp(|\Delta m|) \approx 1$ for close pairs, leaving short-range merges largely unaffected while long-range distances are inflated. Whence, there is an earlier effective connectivity, particularly pronounced for the cluster pattern where within-cluster mark homogeneity is strongest. Under negatively correlated marks the red band shifts strongly to the right and downward. The $4\times 4$ grid assigns values in $\{0, 8\}$ to neighbouring cells by which virtually all nearest-neighbour pairs satisfy $|\Delta m| \approx 8$. This yields $\exp(|\Delta m|) \approx 2981$ and massively delaying all short-range merges. The red band is consequently narrow and confined to large $\varepsilon$, a pattern that is consistent across all three spatial configurations.

The separation between the blue and red bands encodes the degree of mark-space coupling. When marks are random the two bands nearly coincide, reflecting the absence of coupling. When marks are structured the bands diverge, with the direction and magnitude of divergence determined by the interplay between spatial and mark processes. For the cluster pattern with positively correlated marks both bands shift in the same direction (upward), so the mark-weighted ECC is even more extreme than the plain one. For the cluster pattern with negatively correlated marks the two bands shift in opposite directions, producing the largest separation observed in the figure. These results demonstrate that the mark-weighted ECC is sensitive to both the sign and the spatial scale of mark dependence, and that this sensitivity is preserved across qualitatively different point process geometries.
  
\section{Application}

\subsection{Duke forest data}

\begin{figure}[htbp]
    \centering
    \includegraphics[scale=.45]{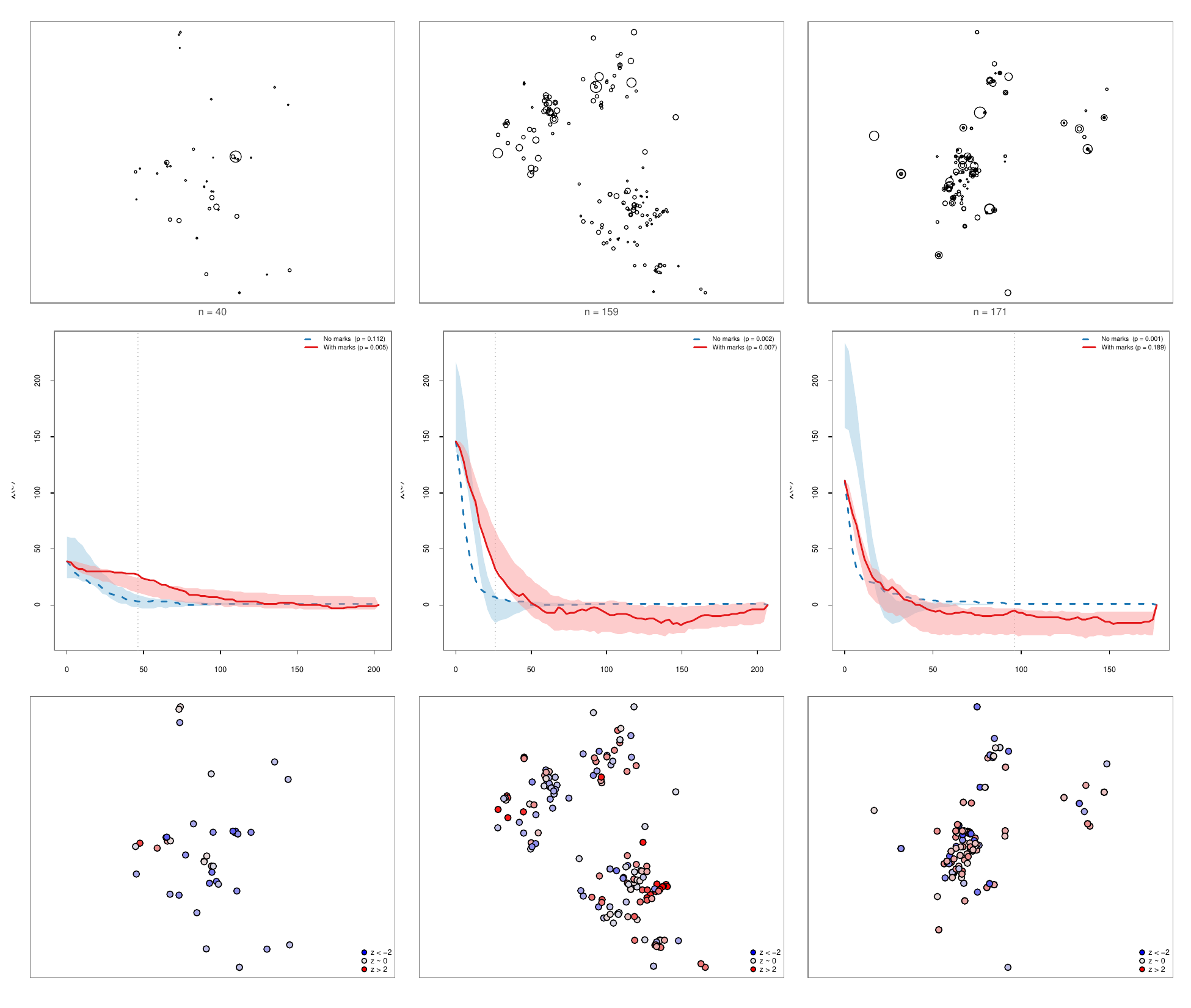}
    \caption{Results for common persimmon, eastern rudbud and American hornbeam from with diameter at breast height as marks left to right. Point patterns of selected species of the Duke forest with diameter of disc corresponding to the mark value (top row), Euler characteristic curves and envelopes computed over 999 simulation for unmarked (blue) and marked (red) point patterns (central row), and computed $Z$-values (bottom row)}
    \label{fig:applDuke}
\end{figure}

The Duke forest data reports the location and the diameter at breast height for 37 distinct tree species measured at some time during 2014. The measurement were collected in an area of size $65{\text{km}}^2$ (convex hull of the locations), which is split into three sub-areas: west, east, and south.  From the 37 tree species, we selected common persimmon (n = 40), eastern rudbud (n = 159) and American hornbeam (n = 171) and computed both the ECC and corresponding envelopes based on $nsim = 999$ simulations. The results are shown in 
Figure \ref{fig:applDuke}.

For the common persimmon (left) the ECC detected no specific structure for the points (p = 0.112) but for the marked points (p = 0.005) with significant deviations from the top of the red envelopes at around $\epsilon = 50$. This indicated that the marks of nearby by are dissimilar such that the points are connected slower as under the independent mark scenario. In contrast, both the unmarked and the marked pattern of the eastern rudbud show signifiant geometric structured ($p=0.002$ and  $p=0.007$) with deviations from both envelopes at short distances up to around $\epsilon = 40$. This indicated that the point appear to be clustered and merge faster as under the CSR setting. At the same time, the marks of nearby points are also more similar in DBH values such that they are connected faster as under the iid setting. 
Finally, for the American hornbeam only the points ($p=0.001$) but not the marks  ($p=0.189$)
reflect significant geometric structure at short distances up to $\epsilon=20$. This indicates that the points show clustering which yields an increase in the Betti0 number.  

\subsection{Anemolia data}

\begin{figure}[htbp]
    \centering
    \includegraphics[scale=.4]{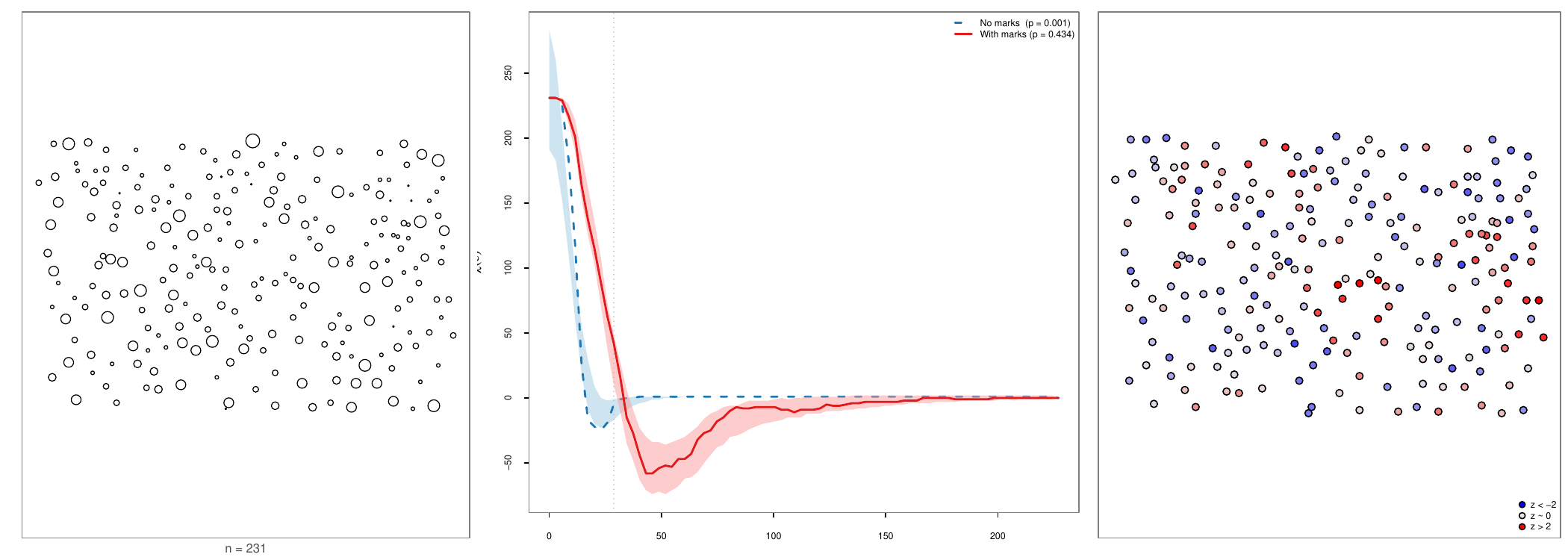}
    \caption{Results for anemones data. Point patterns with diameter of disc corresponding to the mark value (top row), Euler characteristic curves and envelopes computed over 999 simulation for unmarked (blue) and marked (red) point patterns (central row), and computed $Z$-values (bottom row)}
    \label{fig:applanemones}
\end{figure}

The anemones dataset, originally described by \cite{Kooijman1979}  records the spatial locations and diameters of $n = 231$ beadlet sea anemones (Actinia equina) observed on a boulder near sea level. The observation window is a rectangle of $280 \times 180$ units, where one unit corresponds approximately to $0.475$\,cm, giving a physical plot size of roughly $133 \times 85$\,cm. The marks are integer-valued diameters ranging from 1 to 6 units, reflecting the body size of each individual anemone at the time of observation. Inspecting  Figure \ref{fig:applanemones}, the ECC curves (central panel) suggest a significant geometric structure only for the points ($p= 0.001$) while the ECC of the marked points remain completely in the envelopes ($p=0.434$). For the points, the ECC show a clear deviation to the right at very small scales of $\epsilon$ which turns into an deviation to the left as $\epsilon$ increases. THis means that the points merge intially slower  but then faster as expected under CSR. Noting that the blue ECC becomes also negative, Betti1 counting the number of loops becomes larger then Betti0 which reflects the number of connected components.   

\section{Discussion}

The proposed framework represents a significant advancement in the analysis of marked point processes by bridging the gap between attribute-weighted geometry and formal topological inference. Unlike traditional second-order statistics that treat marks as secondary information on fixed Euclidean distances, our method redefines the underlying metric space itself. By employing a mark-weighted distance function, we modulate connectivity directly through attribute similarity, effectively transforming topological structures into a reflection of \textit{attribute-neighborhoods}.

The simulation study provides robust evidence for the framework's discriminative power and specificity. Scenarios 1, 2, and 5 demonstrate that the method is resilient to false positives arising from purely geometric clustering or inhibition, as the Euler Characteristic Curve  remains within the global envelope when marks are independent. Conversely, Scenarios 3, 4, and 6 confirm its high sensitivity to attribute-space coupling: the observed topological collapse in the ECC plots reflects an accelerated merger of components driven by homophily. 

A key innovation is the introduction of local $Z$-score decomposition at the critical scale $\epsilon_{crit}$, which allows for the precise localization of structural drivers. The simulation results confirm that this approach reliably identifies structural hubs and topological barriers, providing a granular diagnostic capability that classical summary statistics lack. By integrating the ECC with the Global Envelope Test framework, we move from purely descriptive topology to rigorous statistical hypothesis testing, offering a comprehensive toolkit for identifying, quantifying, and localizing complex dependencies in marked spatial data.
 
\bibliographystyle{chicago}
\bibliography{TDAmain}

\end{document}